\documentclass{article}

\usepackage{arxiv}

\usepackage[utf8]{inputenc} 
\usepackage[T1]{fontenc}    
\usepackage[breaklinks=true,colorlinks=true,linkcolor=blue,urlcolor=blue,citecolor=blue]{hyperref}
\usepackage{url}            
\usepackage{booktabs}       
\usepackage{amsfonts} 
\usepackage{physics}      
\usepackage{nicefrac}       
\usepackage{microtype}      
\usepackage{lipsum}
\usepackage{graphicx}
\usepackage{upgreek}
\usepackage{cancel}
\usepackage{soul}
\title{Single ion thermal wave packet analyzed via time-of-flight detection}

\author{
Felix Stopp, Luis Ortiz-Guti\'errez, Henri Lehec and Ferdinand Schmidt-Kaler\\
  QUANTUM, Institut f\"ur Physik, Universit\"at Mainz, Staudingerweg 7, 55128 Mainz, Germany\\
  \texttt{felstopp@uni-mainz.de} \\
}

\begin{document}
\maketitle
\begin{abstract}
A single $^{40}$Ca ion is confine in the harmonic potential of a Paul trap and cooled to a temperature of a few mK, with a wave packet of sub-$\upmu$m spatial and sub-$\nicefrac{\text{m}}{\text{s}}$ velocity uncertainty. Deterministically extracted from the Paul trap, the single ion is propagating over a distance of 0.27~m and detected. By engineering the ion extraction process on the initial wave packet, theoretically modeling the ion trajectories, and studying experimentally the time-of-flight distribution, we directly infer the state of the previously trapped ion. This analysis allows for accurate remote sensing of the previous motional excitation in the trap potential, both coherently or incoherently. Our  method paves a way to extract, manipulate and design quantum wave packets also outside of the Paul trap. 
\end{abstract}


\section{Introduction} \label{Sec:Intro}
Trapped charged particles in Paul or Penning traps, or equivalently neutral atoms in magnetic or optical potentials, allow today for outstanding control of the wave packet and its motional state in the harmonic confinement potential~\cite{LEI03,WIN79}. For ions  and for trapped atoms, even the ground state of motion can be reached, where the wave packet is shaped as  a Gaussian and where the momentum and position uncertainty are limited by the Heisenberg uncertainty relation.
Single electrons, protons, antiprotons and highly charged ions also have been transported within segmented Penning traps between spatially separated potential wells to allow for ultra-precise measurement, e.g. measuring g-factors, nuclear properties or testing fundamental symmetries of nature~\cite{HAF03,SMO17,GLA19}. Propagation of electron beams in guiding electric oscillating fields~\cite{HOF11} and also cold atom matter wave shuttles in magnetic guides are investigated for novel sensing and interferometric applications~\cite{SCH05,BER16,PAN19}. Likewise, the fast transport of cold trapped ions in the guiding radio-frequency field of a segmented Paul trap has been successfully demonstrated~\cite{WAL12,BOW12} and is currently used for scalable quantum processing~\cite{KAU20,LEK17}.  In all those cases, a guiding field and a confining three-dimensional potential was employed for the transport.
On the other hand, freely propagating ion and electron beams have many applications in technology and science. This includes microscopy, material science, nano-fabrication by focused ion beam milling and electron lithography. The beams are accelerated, steered and focused by electric and magnetic fields. For specific setups with electron beams~\cite{HAS09}, also for ultra cold atoms~\cite{BLO99,BOL14,MEW97}, even genuine matter-wave properties such as interference have been demonstrated. The release of trapped atoms and observation of arrival times after a free fall is since the beginning used to reveal their temperature~\cite{LETT88, BRZOZ2002}. Neutral atoms may be ionized and subsequently accelerated by electrical fields~\cite{LOPEZ2019}, or electrostatic multi-reflection devices are used for accurate mass spectroscopy or lifetime measurements~\cite{ZAJFMAN2004,ITO2013,ITO2013a,FISCHER2019}. Extraction of ions from Paul traps was demonstrated for single ions~\cite{SCHNITZ2009,IZAWA2010,JACOB2016,GROOT2019}. 

It is the focus of our work here to extend the excellent control over trapped ion wave packets into a situation of free-space propagation. Using a single ion in a Paul trap and optimizing the extraction event out of the potential, we  are investigating the wave packet emission and transport in free space, actually over a distance more than six orders of magnitude larger as compared to its wave packet size. From the time-of-flight signal at the arrival we analyze an initially prepared motional state. In this publication, we start with a description of the experimental setup, we theoretically describe the properties of wave packets, of their motional excitation and free-space propagation. Then, we move over to the extraction mechanism for coherent states and thermal states and discuss how to find their characteristic signature in the time-of-flight measurement. Finally, we sketch applications when extending the method for wave packet propagation in the quantum regime.

\section{Experimental setup} \label{sec:SetUp}

\begin{figure}[t!]
\includegraphics[width=1\textwidth]{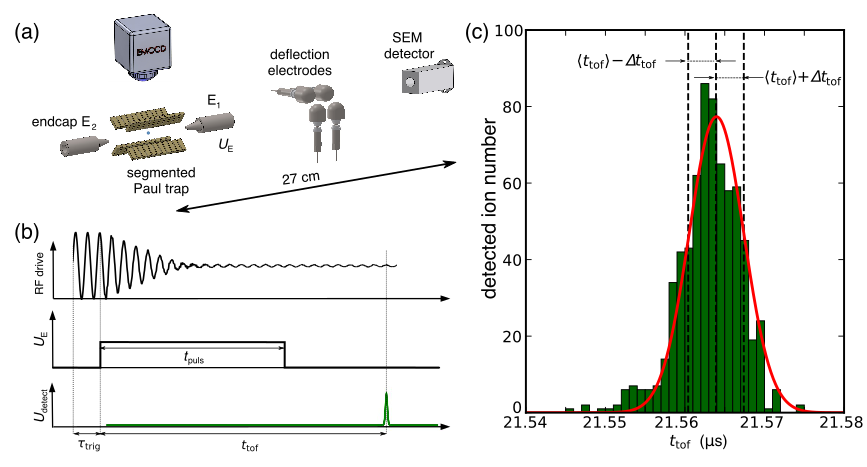}
\caption{\label{Scheme} \textbf{(a)} Scheme of the experimental setup. Ion trap electrodes serve to hold a single ion, laser cooled and emitting fluorescence that is observed by a electron-multiplier-charged-coupled-device (EMCCD) camera. The distance between the center of the trap and the detector is 0.27~m. \textbf{(b)} Time sequence for ion extraction: after a chosen time  delay $\tau_\mathrm{trig}$, triggered on a fixed phase of the RF drive (upper plot), the ion is extracted by applying a square voltage pulse of amplitude $U_\mathrm{E}$ and duration $t_\text{puls}$ (middle plot) to the endcap electrode E$_1$. Synchronously the RF drive starts shutting down. After a delay  $t_\text{tof}$ from the beginning of the $U_\mathrm{E}$ pulse, the extracted ion arrives at the detector and is counted (lower plot). \textbf{(c)} TOF histogram measured after $N=1000$ single ion extraction events at $U_\mathrm{E}= -40\,\text{V}$. $\varDelta t_\text{tof}$ denotes the standard deviation of the Gaussian distribution that we fit to the data.}
\end{figure}

The experimental setup is based on a X-shape segmented Paul trap, see \hyperref[Scheme]{Fig.~\ref{Scheme}(a)}, with geometry similar to that in Ref. \cite{WOL16}. The trap features two chips with RF electrodes  for a radial confinement of $\nicefrac{\omega_\text{rad}}{{2\uppi}} \approx 0.5\,\text{MHz}$ and two chips with DC electrodes, as well as two endcaps E$_1$ and E$_2$, used for extraction and ion state preparation, respectively. The axial trap frequency $\nicefrac{\omega_{z}}{{2\uppi}}$ is controlled between $100\,\text{kHz}$ and $250\,\text{kHz}$ by applying a DC control voltage on the middle segment of the DC chips. Ca ions are produced by photoionization from a neutral atomic beam, trapped and laser cooled on the $4^2 \text{S}_{\nicefrac{1}{2}}$ - $4^2 \text{P}_{\nicefrac{1}{2}}$  dipole transition near 397 nm and a laser near 866 nm is used to pump out of the metastable $3{}^2  \text{D}_{\nicefrac{5}{2}}$ level. In this way, continuous laser-induced fluorescence is emitted near $397\,\text{nm}$. We image this light through an objective on an EMCCD camera, with a $11.55(13)$ magnification, which allows us to resolve the number of ions in the crystal. All experiments in this work have been performed with a single ion. 

The thermal state after Doppler-cooling can be excited to a higher average thermal phonon number by controlled heating of the ion. In such a way, the Gaussian width of the distribution increases both in position and momentum. Alternatively, we can engineer a coherently-excited thermal state by applying an oscillating electrical potential at $\omega_z$ on the endcap E$_2$, with the magnitude of the coherent excitation being controlled by the amplitude of the applied voltage. After initialization, the single ion is extracted by switching on a voltage of $-40\,\text{V}$ to endcap E$_1$ which is pierced with a $400\,\upmu\text{m}$ hole and located 1.45 mm from the trap center. Extracted by the  electric field, the ion is passing through the endcap hole, further steered by several deflection electrodes and finally detected by a secondary electron multiplier, 272 mm downstream. We repeat this procedure of loading, initialization and extraction, for a number of times $N$ and build up a time-of-flight (TOF) histogram, see \hyperref[Scheme]{Fig.~\ref{Scheme}(c)}. 

\begin{figure}[t!]
\includegraphics[width=1\textwidth]{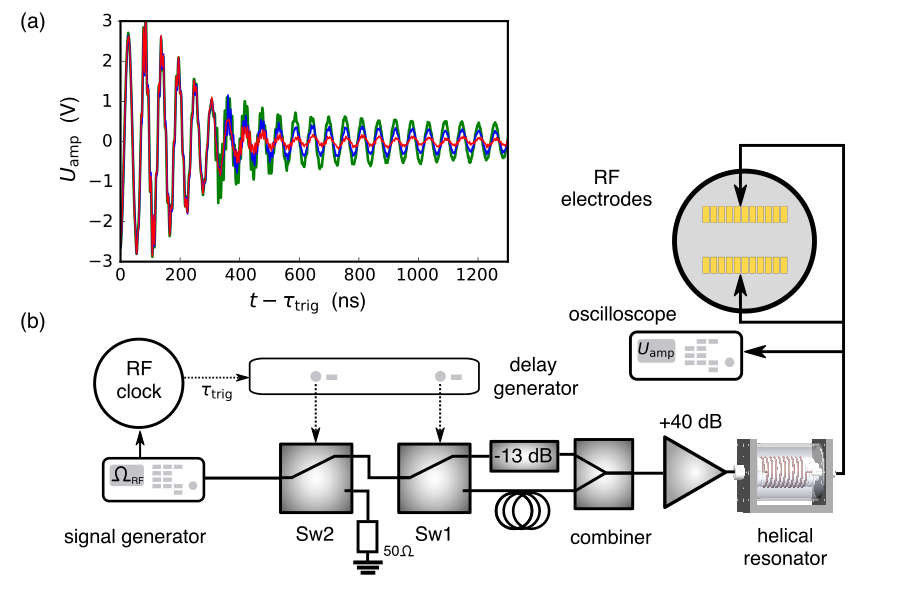}
\caption{\label{rf switching} Dynamic control of the trap drive RF power: \textbf{(a)} Examples of traces obtained with the oscilloscope are shown (red, blue, green). After a fast initial down-ramping, the level of remaining RF power is chosen by a proper timing of the RF-switching. \textbf{(b)} The RF setup consists of a signal generator Keithley-3390, whose output passes a series of two RF switches Sw(1,2), each  controlled by one of the TTL outputs of a SRS-DG535 delay generator. The top RF-branch has a fixed attenuation, while second RF-branch has zero attenuation and a relative phase shift of $\uppi$. Initially using the top branch we switch with Sw1 to the lower branch and use the destructive interference from the strong RF-branch to quickly reduce the power in the helical resonator, much faster as compared to its decay time of $\sim 2\,\upmu$s. Near the time when the helical resonator is almost empty, we employ switching Sw2 to control the level of the remaining RF.}
\end{figure}

In order to optimize the single ion beam performance at low extraction energy, i.e. switching on a DC voltage of $U_\text{E}=-40\,\text{V}$ for $t_\text{puls}=1600\,\text{ns}$, special attention has been paid to the extraction time sequence, see \hyperref[Scheme]{Fig.~\ref{Scheme}(b)}.

Depending on the amplitude and phase of the trap RF drive fringe field component along the trap axis, the ion can either be accelerated or slowed down in the vicinity of the endcap E$_1$, causing in turn a modification of the TOF distribution. In order to have full control over these effects, the extraction DC voltage trigger is synchronized, using a delay generator with arbitrary phase with respect to the phase of the RF drive. Also, the RF drive amplitude is switched synchronously with the extraction event. With a typical switch-off time of $\sim$350~ns, the remaining RF amplitude has already been reduced to the residual level and a corresponding electric fringe field is present in vicinity of the endcap when the ion reaches this position. Through a fine tuning of the temporal shaping of the RF amplitude and phase, ion wave packets at arbitrary extraction conditions can be realized. The control of the RF drive is described in \hyperref[rf switching]{Fig.~\ref{rf switching}(a-b)}: choosing the RF switching phase around $\varphi_\text{RF}=0$ and optimizing the residual RF amplitude during extraction, we can tailor the electric forces due to the RF fringe field, when the wave packet passes the endcap E$_1$. Note, that the initial ramping-down phase is fast as compared to the ion motion at low extraction DC voltages, such that the forces on the wave packet are indeed controlled by the freely adjustable residual and constant RF amplitude. We will show, that by tuning its amplitude and phase, the wave packet is either narrowed down or stretched out during the extraction event. The latter setting allows for an improved sensitivity of the TOF distribution on details of the previously prepared motional state of the wave packet, i.e. the RF residual amplitude may act as a magnification tool.

For a better understanding of the ion dynamics inside the trap and during the extraction process, we will revisit in the next section some theoretical elements for driven harmonic oscillator wave packets.  We will provide tools to calculate TOF distributions associated with the extraction of coherently or incoherently excited states.

\section{Elements of theory for a trapped and for an accelerated ion wave packet}
\label{theory} 

\subsection{Modeling the ion TOF distribution for a thermal wave packet}
\begin{figure}[t!]
\includegraphics[width=1\textwidth]{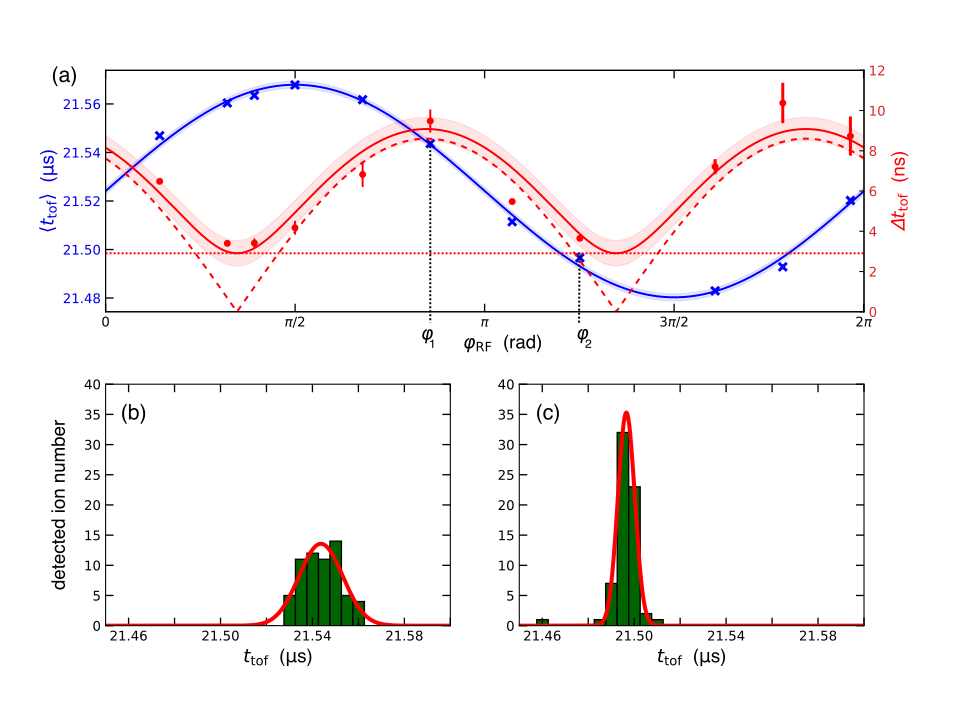}
\caption{\textbf{(a)} Measured TOF histogram: The average flight time $\langle t_\mathrm{tof}\rangle$ and the width of the distribution $\varDelta t_\text{tof}$ are depending on the initial wave packet with $\omega_z=2\uppi\cdot 150\,\text{kHz}$ and the RF drive phase $\varphi_\text{RF}$ at which the ion was extracted with $U_\mathrm{E}=-40\,\text{V}$. The average is modulated by a sine-function (blue) with a frequency of $\Omega_\mathrm{RF}=2\uppi\cdot 17.84\,\text{MHz}$ due to the electrical forces from the residual RF field. The width  $\varDelta t_{\text{tof}}$ follows the squared sum (red) of a technical noise contribution $\varDelta t_{\text{tech}}$ (red dotted) and of the effects due to the RF forces $\varDelta t_{\text{mod}}$ (red dashed), see \hyperref[Deltaaaaa]{eq.~\ref{Deltaaaaa}} and text for details. Shaded regions indicate the 1-$\sigma$ margin of the fits. \textbf{(b-c)} Examples for measured TOF distributions demonstrating the variation of the average value and width, using different extraction phases $\varphi_\text{RF}$. Each histogram is taken with 100 single ion extractions. At phase $\varphi_1$, see \textbf{(b)}, the width is close to a maximum and at $\varphi_2$, see \textbf{(c)}, the width is close to a minimum.}
\label{TOF40}
\end{figure}

A thermal wave packet in a harmonic potential is a statistical mixture of the oscillator eigenstates $|n\rangle$  with population probabilities $P_n$, given by the distribution of axial phonons $P^\text{th}_n=\frac{\langle n\rangle^n}{(\langle n\rangle+1)^{n+1}}$ where  
\begin{align}
\langle n \rangle= \frac{1}{\exp\left(\frac{\hbar\omega_z}{k_\text{B}T}\right)-1}
\label{phononnum}
\end{align}
denotes the temperature dependent mean phonon number. The time-averaged momentum distribution of the ion takes then the form of a Gaussian~\cite{KNU12}:
\begin{equation}
\Psi^\text{th}(p_z)= \frac{1}{\sqrt{2 \uppi}\varDelta p^\text{th}_z} \exp\Big({-\frac{p_z^2}{2(\varDelta p^\text{th}_z)^2}}\Big),
\end{equation} 
with a root mean square (RMS) width that reads $\varDelta p^\text{th}_z \approx \sqrt{m k_\text{B} T}  $, assuming the weak binding approximation $k_\text{B} T \gg\hbar \omega_z$, which is a good approximation for our operation conditions. Note, that this approximation is well justified for Doppler-cooled ions with axial trap frequencies $\nicefrac{\omega_z}{2\uppi}$ of the order of 100~kHz, which are the typical conditions in this work. Similarly, in position space, the distribution takes a Gaussian form with a RMS width of 
\begin{align}
    \varDelta z^\text{th}\approx \sqrt{\frac{k_\text{B} T}{m\omega_z ^2}}.
\label{deltazth}
\end{align}
After the extraction the initial momentum distribution can be mapped onto the TOF distribution. We approximate the theoretical TOF distribution by a Gaussian probability distribution, defined by its mean value $\langle t_{\text{tof}} \rangle$ and its standard deviation $\varDelta t_{\text{tof}}$, see \hyperref[Scheme]{Fig.~\ref{Scheme}(c)}. The average $\langle t_{\text{tof}} \rangle$ of the TOF can be decomposed into two parts:
\begin{align}
    \langle t_{\text{tof}} \rangle = t_{\text{phys}} + t_{\text{tech}},
\end{align}
where $ t_{\text{tech}} $ accounts for fixed technical delays (cables, switches, detector, ect.) and $t_{\text{phys}}$ results from the physical effects expected for an ion, initially at rest at the trap center and submitted to an electric extraction potential $U_\text{E}$, followed by a free flight into the detector. In our setup, a significant axial field contribution from the residual RF drive is present in front of the endcap and adds to the extraction potential $U_\text{E}$. The physical TOF $t_{\text{phys}}$ can be modeled as

\begin{equation}
\label{tphys}
t_{\text{phys}}= \sqrt{\frac{md^2}{2 e \kappa U_\text{E}}}+t_\text{mod}\sin\underbrace{(\Omega_\text{RF} \tau_\text{trig} +\phi_0)}_{:=\varphi_\text{RF}},
\end{equation}
where the first term is the physical TOF from the trap into the detector at distance $d$, without the contribution of the RF drive. For the acceleration process we take the geometry factor $\kappa$ into account. The second term is the contribution by the RF drive field. The time  $t_\text{mod}$ denotes the amplitude of the TOF modulation and depends on the magnitude of the RF residual field. The behaviour of \hyperref[tphys]{eq.~\ref{tphys}} is experimentally very well verified by our measurements as shown in \hyperref[TOF40]{Fig.~\ref{TOF40}(a-c)}, where we find a sinusoidal modulation. 

\begin{figure}
\centering
\includegraphics[width=1\textwidth]{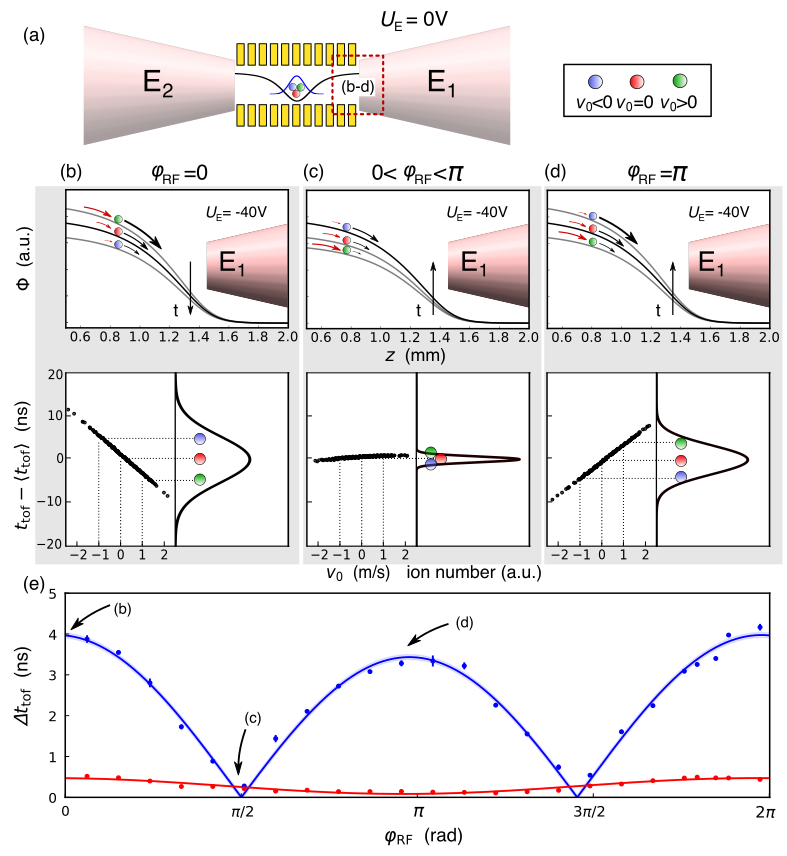}
\caption{\textbf{(a)} Schematic representation of a thermal Gaussian wave packet (blue curve, not to scale) in the trap potential between both endcaps E$_1$ and E$_2$ (black curve). Ions have different initial velocities within the momentum distribution: $v_0=0$ (red), $v_0>0$ (green) and $v_0<0$ (blue). As soon as the extraction starts, the ions pass through different potentials, caused by the periodic RF potential additional to the extraction potential in the red dashed area. \textbf{(b)} Temporal potential curve and final TOF distribution for $\varphi_\text{RF}=0$: Initial ion with $v_0=0$ (red) is not influenced by the RF potential, initial ion with $v_0>0$ (green) experiences a higher gradient, ion with $v_0<0$ (blue) a lower gradient. This leads to a broadening of the distribution. \textbf{(c)} Temporal potential curve and final TOF distribution for $0<\varphi_\text{RF}<\uppi$: 
The phase is precisely chosen such that initially slower ions perceive a higher gradient. All ions reach the detector at the same time. \textbf{(d)} Temporal potential curve and final TOF distribution for $\varphi_\text{RF}=\uppi$: Initially slower ions experience a stronger gradient and overtake the initially faster ions. \textbf{(e)} Numerically determined Gaussian distribution for $T=3\,\text{mK}$ for an exponentially decreasing RF field with $U_\text{RF}=10\,\text{V}$ (blue, focus between trap and detector) and $U_\text{RF}=1\,\text{V}$ (red, focus at the detector's position). Each simulation point performed with 300 ions.}
\label{Potential}
\end{figure}

In order to understand the wave packet-stretching effect, we will now investigate the ion velocity during the acceleration phase between the center of the trap and endcap: Before application of the extraction potential, the velocity distribution follows a Gaussian form, outlined in \hyperref[TOF40]{Fig.~\ref{Potential}(a)}. As soon the extraction potential $\Phi_\text{E}(z)$ is applied, ions with different initial velocities, marked with red ($v_0=0$), green ($v_0>0$) and blue ($v_0<0$), are accelerated. The remaining RF field causes an additional periodic potential $\Phi_\text{RF}(z,t)$ that increases or decreases the gradient along the $z$ axis. The overall potential can thus be written as $\Phi(z,t)=\Phi_\text{E}(z)+\Phi_\text{RF}(z)\sin(\Omega_\text{RF}t)$.
The potential $\Phi(z,t)$ in front of endcap E$_1$ is shown in \hyperref[Potential]{Fig.~\ref{Potential}(b-d)}. Here the bare extraction potential, caused by the voltage $U_\text{E}=-40\,\text{V}$, is shown in black 
and the total potential which contains the periodic contribution is represented in grey. The strongest force acting on the ion, given by the potential gradient, is at $z_\text{grad}\simeq1.2\,\text{mm}$. We investigate the interaction on the ions at this position more closely and distinguish three cases. \hyperref[Potential]{Fig.~4(b)}: The extraction is started at phase $\varphi_\text{RF}=0$. Ions without an initial momentum (red) pass through the unmodified potential at $z_\text{grad}$. In the ion momentum distribution, events with higher velocity (green) are more accelerated by the higher gradient, which was there a short moment before, while slower velocities (blue) are less accelerated. The initial Gaussian distribution is thereby stretched further symmetrically. The situation is inverted for the setting $\varphi_\text{RF}=\uppi$, described in \hyperref[Potential]{Fig.~4(d)}.
Ions initially at higher velocities of the distribution are decelerated by the RF potential, while slower ions are accelerated. Depending on the potential strengths $\Phi_\text{E}$ and $\Phi_\text{RF}$, initially slower ions may overtake the faster ones and reach the detector first and cause again a wider flight time distribution. This process is acts symmetrical, such that a Gaussian shape is conserved. Adjusting parameters, the distribution may be focused on any position $z_\text{foc}$. This point can be shifted to the detector position $z_\text{foc} =z_\text{det}$ by lowering the RF potential. If the point is still between the trap and the detector, $0<z_\text{foc} <z_\text{det}$, then there is a phase $0<\varphi_\text{RF}<\uppi$ focusing is achieved approximately on the detector, see \hyperref[Potential]{Fig.~4(c)}. Under typical experimental conditions, however, the spread of the wave packet is small as compared to the RF period, such that we can approximate the TOF distribution by a Gaussian and only in cases of extreme compression becomes asymmetric. This is due to the fact, that the maximum modulation is $t_\text{mod}$.

To fully model the experimental data, we include an additional broadening  $\varDelta t_{\text{tech}}$ into account, independent from $\varphi_\text{RF}$. This constant term accounts for jitter in the switches, in the delay generator triggers and also jitter in the SEM ion detector and counting. As a result, the width of the TOF distribution $\varDelta t_{\text{tof}}$ now reads
\begin{equation}
\label{Deltaaaaa}
 \varDelta t_{\text{tof}} = \sqrt{ \varDelta t_{\text{phys}}^2+\varDelta t_{\text{tech}}^2}.
\end{equation}

The predicted periodic change of the width $\varDelta t_\text{tof}$ as a function of the extraction phase $\varphi_\text{RF}$ is fully confirmed by our measurements, see \hyperref[TOF40]{Fig.~\ref{TOF40}(a)}. For these experimental runs, technical limitations set a lower limit of the TOF widths, fitted here to $\varDelta t_{\text{tech}} = 2.9(7)\,\text{ns}$, dominated by the intrinsic width of the detected ion signal. Indeed, the typical width of a single ion voltage drop pulse in our setup is  $2\,\text{ns}$. Additionally, a trigger jitter of around $0.5\,\text{ns}$ leads to further technical broadening. However, we observe a phase difference between maximum average shift at $\nicefrac{\uppi}{2}$ and maximum wave packet compression of about $0.5\,\text{rad}$. We conjecture that this is caused by deviations of the DC extraction voltage $U_\text{E}$ from a perfect square pulse. Indeed, we observe a ramping in the order of $100\,\text{ns}$, as the switch and the supply have to load the endcap, which acts like a capacitor. This unsymmetrical DC waveform may cause additional compressing effects in the ion acceleration.

Since the width of the TOF distribution $\varDelta t_\text{phys}$ is proportional to the initial momentum distribution width, we model the TOF distribution at $\varphi_\text{RF}=0$ for a thermal state as $\varDelta t^\text{th}_{\text{phys}} \sim  \sqrt{T}$, where $T$ denotes the temperature and the proportionality factor depends on a combination of $d$, $q$, $m$, $U_\text{RF}$ and $U_\text{E}$. It is the goal of this work to optimize the wave packet stretching such that TOF measurements are sufficiently sensitive to reveal the temperature of an ion in the trap, see \hyperref[thermal]{subsect.~\ref{thermal}}. Also, we will employ TOF measurements to determine the heating of the ion, see \hyperref[heating]{subsect.~\ref{heating}}.

\subsection{Ion TOF distribution for a coherently excited thermal wave packet}

To coherently excite a wave packet, we apply an electric sinusoidal drive resonant with the trapping frequency $\nicefrac{\omega_z}{2\uppi}$, coined \textit{tickling}, thus generating a momentum modulation. Therefore, the timing at which the ion arrives in the vicinity of E$_1$ will be modulated. The phase of the coherent oscillation $\phi^\text{coh}$ and its amplitude $\varphi_\text{mod}$ can be set at arbitrary values by the experimental parameters of the tickling. We describe the effect as an additional oscillating phase inside the modulation of \hyperref[tphys]{eq.~\ref{tphys}}. The overall phase is then transformed to
\begin{equation}
\label{phitilde}
\varphi^\text{coh}_\text{RF}=\varphi_\text{RF}+\varphi_\text{mod}\sin(\phi^\text{coh}).
\end{equation}

Similarly to the previously discussed case for thermal wave packets, we find the average of the TOF distribution and its width modified by the coherent excitation. For a phase of $\varphi_\text{RF}=0$, the linear magnification as compared to the initial $t_\text{mod}$ results in a TOF characterized by an average value
\begin{equation}
     \langle t_{\text{tof}} \rangle =  \sqrt{\frac{md^2}{2 e \kappa U_\text{E}}}+t_\text{mod}\sin[\varphi_\text{mod}\sin(\phi^\text{coh})] + t_{\text{tech}}.
\label{TOFCoh1}
\end{equation}
Depending on the chosen setting of the phase $\phi^\text{coh}$ of the coherent amplitude, the minimum and maximum average TOF are realized by
\begin{equation}
     \langle t^\text{max/min}_{\text{tof}} \rangle =  \sqrt{\frac{md^2}{2 e \kappa U_\text{E}}}\pm t_\text{mod}\sin(\varphi_\text{mod}) + t_{\text{tech}}.
\label{TOFCoh2}
\end{equation}

Again, we included the technical contribution to fully model the experimental situation. Experiments with single ion extraction in coherent states will be further explored in \hyperref[coherent]{sect.~\ref{coherent}}. 

\section{Experimental implementation of coherent wave packet excitation and characterization of the wave packet-stretching mechanism}
\label{coherent}
\begin{figure}[t!]
\centering
\includegraphics[width=1\textwidth]{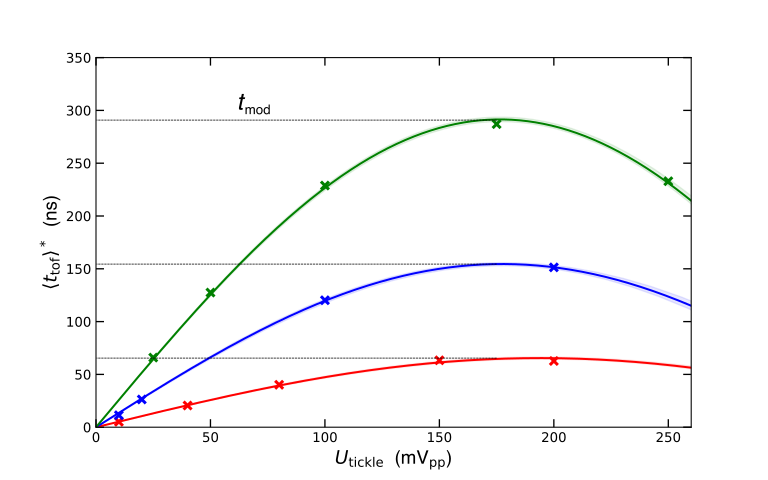}
\caption{Measured TOF shift $\langle t_\mathrm{tof}\rangle^*$ for 
corresponding $t_\text{mod}$ of $65(1)\,\text{ns}$ (red), $155(2)\,\text{ns}$ (blue) and $291(2)\,\text{ns}$ (green).
Depending on the residual RF amplitude the resonant coherent excitation characterized by $U_\mathrm{tickle}$ results in a TOF variation. We fit the prediction of \hyperref[tofdiff]{eq.~\ref{tofdiff}}. Shaded areas show 1-$\sigma$ margins. }
\label{Coherent2}
\end{figure}
We extract coherent states to study the stretching mechanism experimentally. Herefore, we excite the single trapped ion by the sinusoidal drive resonant to the axial oscillation frequency $\nicefrac{\omega_z}{2\uppi}=247\,\text{kHz}$  \cite{Carruthers,Jefferts,Leibfried}, in our case applied to endcap E$_2$, see \hyperref[Scheme]{Fig.~\ref{Scheme}(a)}. Note, that before and during the excitation, the ion is exposed to Doppler-cooling. We then make sure that a steady state is reached. We denote this convoluted wave packet as thermal-coherent state, however, under typical experimental conditions the coherent amplitude exceeds the thermal one by far.

Single ion extraction events are characterized by the residual RF amplitude and the extraction phase $\varphi_\text{mod}$, which is controlled by the time delay of the DC square pulse. Additionally, the phase of the coherent drive $\phi^\text{coh}$ and its amplitude $U_\text{tickle}$ can be adjusted to determine the average position and momentum of the wave packet at the extraction instant. As seen from \hyperref[TOFCoh1]{eq.~\ref{TOFCoh1}} we have a sine function as the argument of another sine function, therefore we could, in principle, have very rich structures on the TOF mapping. Hereafter, we will treat $\langle t_\text{tof} \rangle$ as a function of one variable, the coherent excitation phase $\phi^\text{coh}$. In the low $\varphi_\text{mod}$ regime, $\langle t_\text{tof}\rangle$ is a small amplitude sinusoid. However for higher coherent excitation amplitudes it turns into a sine-enveloped sine function, where the maxima and minima start to approach the central value after reaching $\pm t_\text{mod}$. 

Experimentally, we set this $t_\text{mod}$ value with the remaining RF amplitude after the partial cancellation in front of the pierced endcap E$_1$. It corresponds to our magnifying factor and can be tuned by means of a delay generator on the trigger time $\tau_\text{trig}$ when switching the RF signal off. This mechanism works by taking advantage of an extra momentum obtained by to the residual RF at the extraction moment. 
It is of crucial importance in order to be able to time-resolve the relatively small in-trap momenta via TOF detection method. 

In order to characterize and quantify this magnification, we identify the two coherent phases $\phi^\text{coh}$ at which a maximum average TOF and a minimum average TOF take place, $\langle t_\text{tof}^\text{max}\rangle$ and $\langle t_\text{tof}^\text{min}\rangle$ respectively. Once these coherent phases are identified and kept fixed, through multiple extractions we can plot the difference between these two extreme average TOF values as a function of the coherent excitation voltage amplitude.

The magnifying mechanism is therefore characterized by repeating this procedure for three different RF-cancellations shown in \hyperref[rf switching]{Fig.~\ref{rf switching}(a)}. Scanning the excitation voltage from $10\,\text{mV}_\text{pp}$ to $250\,\text{mV}_\text{pp}$, at a certain intermediate voltage, $\langle t_\text{tof}^\text{max}\rangle$ and $\langle t_\text{tof}^\text{min}\rangle$ are maximally delayed, however for lower and higher coherent amplitudes this difference starts to decrease as shown in \hyperref[Coherent2]{Fig.~\ref{Coherent2}}. For a very high voltage of the driving coherent excitation, an inverted difference is expected. The prediction (solid lines in Fig.) are given by

\begin{align}
\label{tofdiff}
\langle t_\text{tof}\rangle^*\equiv\frac{\langle t^\text{max}_\text{tof}\rangle-\langle t^\text{min}_\text{tof}\rangle}{2}= t_\text{mod}\cdot \sin(\varphi_\text{mod}).    
\end{align}

All the three cases show the same periodic behaviour $\varphi_\text{mod}=\alpha U_\text{tickle}$ with $\alpha=8.5(5)\nicefrac{\text{mrad}}{\text{mV}_\text{pp}}$, as expected. This measurement gives us a more accurate way of determining the magnification factor of our system, in contrast to the coarse estimate via the degree of cancellation of the RF field.

\section{Mapping the momentum distribution of thermal states from TOF measurements}

 \subsection{Temperature measurements and calibration}

\label{thermal}
\begin{figure}[t!]
\includegraphics[width=1\textwidth]{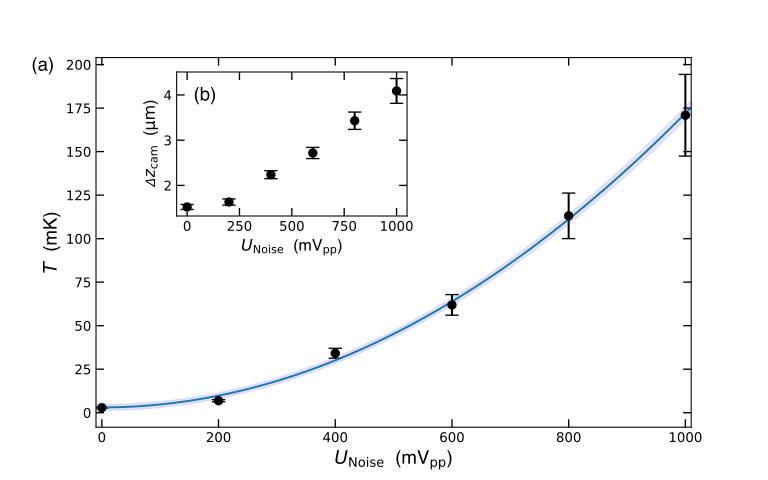}
\caption{\label{TempNoise}\textbf{(a)} Ion size-based  temperatures $T$ against the applied white noise amplitude $U_\mathrm{Noise}$. \textbf{(b)} These temperatures are deduced from the ion size on the camera picture $\varDelta z_\mathrm{cam}$ at an axial trap frequency of $\omega_z=2\uppi\cdot 247\,\text{kHz}$ applied for different noise amplitudes.}
\end{figure}

\begin{figure}[t!]
\includegraphics[width=1\textwidth]{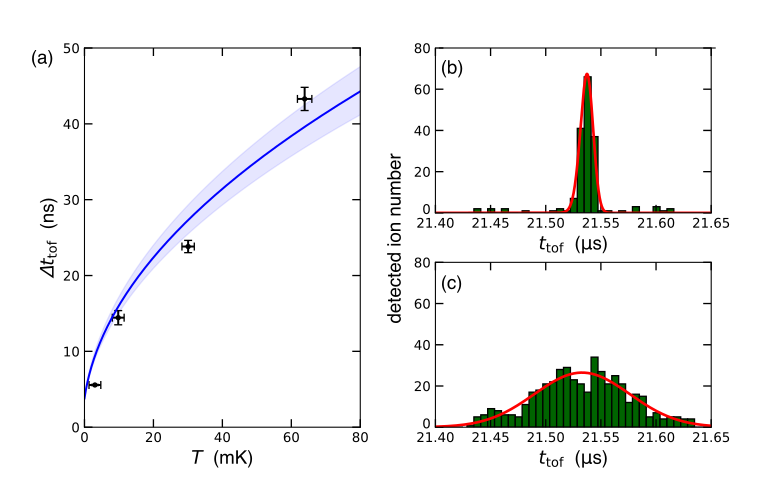}
\caption{\label{TOFTemp}Mapping the momentum distribution of the thermal states. \textbf{(a)} Black points: width of the TOF distribution. Blue line: fit using a square root function.  \textbf{(b)} and \textbf{(c)} Histograms of the TOF distribution at respectively low temperature $T_0\approx 3\,\mathrm{mK}$ and high temperature $T\approx65\,\mathrm{mK}$. From the fits (red curve) we deduce a respective TOF broadening of $\varDelta t_\text{tof}=5.57(8)\,\text{ns}$ and $\varDelta t_\text{tof}=43(2)\,\text{ns}$.}
\end{figure}

We will first characterize our TOF based thermometry technique using ions with arbitrary temperatures. In order to tune the initial temperature of the ions, we apply white noise of arbitrary amplitude $U_{\text{Noise}} $ on the endcap E$_2$. For this geometry, the axial degree of freedom is addressed. Thus, we can tune the temperature associated with the axial motion, typically between about 1~mK up to 200~mK. In order to relate the applied noise amplitude with temperature, we use the spatial thermometry technique~\cite{KNU12,SRI19}. We determine the RMS width of the ion $\varDelta z^{\text{th}}$, see \hyperref[TempNoise]{Fig.~\ref{TempNoise}(b)}, and taking into account the point spread function $\varDelta z_\text{PSF}=1.44(2)\,\upmu\text{m}$ of the imaging system, we infer the temperature, see \hyperref[deltazth]{eq.~\ref{deltazth}}.  This data are shown in \hyperref[TempNoise]{Fig.~\ref{TempNoise}(a)}, featuring the expected quadratic behaviour of the temperature with the noise amplitude.

Let us now focus on the TOF measurement after extraction. After the initialization in the thermal state, the applied noise is stopped simultaneously with the extraction event. The choice of the RF drive phase $\varphi_\text{RF}$ at extraction turns out of particular importance. In order to get a good mapping of the initial velocities on the TOF signal, we chose a RF drive phase centered at the linear regime at $\varphi_\text{RF}\simeq 0$ of the average TOF sinusoidal modulation, see \hyperref[tphys]{eq.~\ref{tphys}}. Second, we chose a phase corresponding to an ascending slope for maximal stretching of the initial momentum distribution, see explanations in \hyperref[theory]{sect.~\ref{theory}}. For the residual RF amplitude, we test the effect in two different settings: at an intermediate stretching (blue curve in \hyperref[Coherent2]{Fig.~\ref{Coherent2}}) at minimal stretching (red). As expected, the latter does not feature a measurable large TOF modification depending on initial temperatures. However, at the intermediate RF amplitude we observe a sufficiently large stretching for temperatures up to  $100\,\text{mK}$. For high amplitudes, the TOF stretching acts in non-linear way in case of large initial momenta.

Results are shown in \hyperref[TOFTemp]{Fig.~\ref{TOFTemp}(a-c)}.  In order to analyse the experimental results, we fit the measured points with the expected function  $\varDelta t_\text{tof}=\sqrt{b\cdot T+\varDelta t_\text{tech}^2}$ where the technical noise $t_\text{tech}$ had been already determinated in previous measurements, see \hyperref[TOF40]{Fig.~\ref{TOF40}(a)}.  We can see a good agreement of this  model with the experimental data.  From the fit, we extract a TOF width dependance in the temperature of $b=24(4)\, \nicefrac{(\text{ns})^2}{\text{mK}}$. It is remarkable that such sensitivity is sufficient to resolve inital momenta of the trapped ion, corresponding to an axial temperature of $3\,\text{mK}$. As discussed before, the sensitivity can be increased when using a higher residual RF amplitude, if lower thermal states should be analized from the TOF signal. The ultimate limit on the wavepacket resolution is resulting from technical noise, and for a realistic value of 1~ns, temperatures of $50\,\upmu\text{K}$ become accessible. This corresponds to sideband cooling of the ion near to the ground state. On the other hand, this technique can be adapted for high ion temperatures, which are hard to determine from sideband spectroscopy~\cite{DIE89} and even by dark resonance spectroscopy~\cite{ROS15}.

\subsection{Determination of the heating rate}
\label{heating}

A first application of the thermometry is the determination of the heating rate. For this, we keep the ion in the dark after laser cooling for a variable time $t_\text{h}$. Only then, we apply the extraction and determine from the TOF distribution the ion temperature increase with $t_\text{h}$. We use the calibration from the previous section for converting TOF distributions into  temperatures. Conviniently, we have chose parameters for a linear stretching. The heating rate of $\dot{\langle n\rangle}=1.9(3) \ \nicefrac{\text{ph}}{\text{ms}}$ is within the error margins in agreement with similar traps in our lab. We emphasize that this new temperature measurement technique does not require laser spectroscopy, even no laser excitation and may be favourable e.g. to study the motional state and the cooling rates for ions of other species or of different isotopes, which are sympathetically cooled by Coulomb interaction from a laser-cooled ion or ion crystal.

\begin{figure}
\centering
\includegraphics[width=1\textwidth]{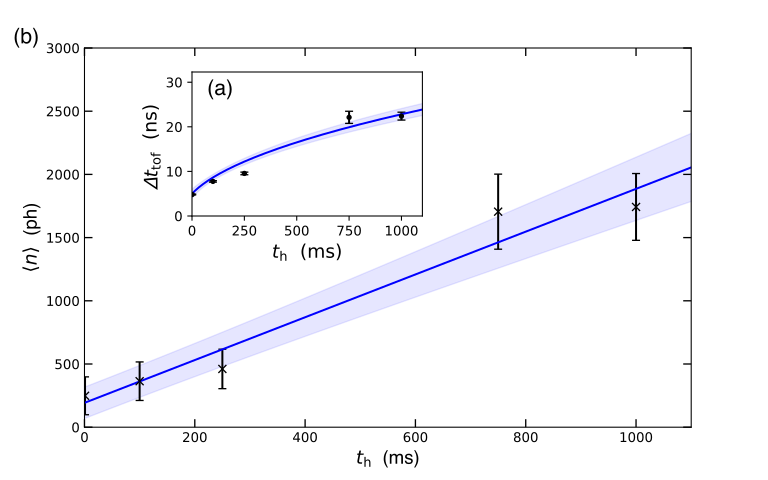}
\caption{\label{TempHeat}\textbf{(a)} Axial broadening of the TOF distribution depending of the time in which the ion is heated up. \textbf{(b)} Corresponding graph, which compares the measurement of the heating time with the ion temperature and the deduced phonon number.}
\end{figure}

\section{Conclusion and Outlook} \label{Sec:Conclusion}

Using a single ion wave packet, we have experimentally demonstrated how to control thermal and coherent excitation of the motion inside the trap, and how a designed extraction out of the harmonic oscillator potential conserves, or even magnifies the characteristic traits of the wave packet. From recording a TOF distribution we unambiguously detect the coherent and the thermal excitation. 

In the future, we intend to cool the ion to near its quantum ground state and extend our methods and investigations into the quantum regime. We aim for expanding and propagating a ground state wave packet \cite{FUER2014} in free space and explore its TOF distribution. Such experiments will contribute to a better understanding of quantum wave packets propagation and may be of fundamental interest, e.g. for investigating the TOF detection statistics as a novel test of quantum mechanics \cite{MAC2020}, possibly also testing Bohmian predictions \cite{DAS2019}. Potentially, one may also extract the ionic wave packet through a tunneling barrier which could be controlled by tightly focused laser beams that induce optical potentials and experimentally record the time that the wave packet has spent in the barrier region \cite{SAI2020}. Such fundamental tests of quantum mechanics have attracted much discussion recently. The extremely well-controlled extracted single ion wave packet may complement experiments with tunneling cold atoms~\cite{RAM2020} or laser ionization~\cite{CAM2017}. The high control over single extracted ions at low energies may be utilized for experiments on friction near surfaces~\cite{IN2019} or for injecting ions into surface structures with strong magnetic gradients to realize the Stern Gerlach effect with charged beams~\cite{HEN2019}.

On the technological side, the long-distance transport of ions in free space may open alternatives for scalable ion-qubit architectures, where ion traps are interconnected via free space segments~\cite{LEK17}, and ion-qubit Bell pairs are distributed for later using this resource via gate teleportation or entanglement swapping \cite{RIE2008}.

\section{Acknowledgements}
This work has been supported by the Deutsche Forschungsgemeinschaft through the DIP program (Schm 1049/7-1) and by the VW Stiftung. We thank Uli Poschinger and Daniel Pijn for careful reading and helpful comments. We remember in this work the vital encouragement by Prof. Dr. Detlef D\"urr, who passed away January 2021, much too early.



\end{document}